\documentclass[11pt]{article}
\usepackage{axodraw,amsmath,epsfig}
\begin{document}

\renewcommand{\theequation}{\thesection.\arabic{equation}}
\newcounter{saveeqn}
\newcommand{\add}{\addtocounter{equation}{1}}
\newcommand{\alpheqn}{\setcounter{saveeqn}{\value{equation}}%
\setcounter{equation}{0}%
\renewcommand{\theequation}{\mbox{\thesection.\arabic{saveeqn}{\alph{equation}}}}}
\newcommand{\reseteqn}{\setcounter{equation}{\value{saveeqn}}%
\renewcommand{\theequation}{\thesection.\arabic{equation}}}
\newenvironment{nedalph}{\add\alpheqn\begin{eqnarray}}{\end{eqnarray}\reseteqn}
%%%%%%%%%%%%%%%%%%%%%%%%%%%%%%%%%%%%%%%%%%%%%%%%%%%%%%%%%5
%%%%%%     pSLASH
\newsavebox{\PSLASH}
\sbox{\PSLASH}{$p$\hspace{-1.8mm}/}
\newcommand{\PS}{\usebox{\PSLASH}}
%%%%%%%%%%%%%%%%%%%%%%%%%%%%%%%%%%%%%%%%%%%%%%%%%%%%%%%%%5
%%%%%%     partialSLASH
\newsavebox{\PARTIALSLASH}
\sbox{\PARTIALSLASH}{$\partial$\hspace{-2.3mm}/}
\newcommand{\PARTIALS}{\usebox{\PARTIALSLASH}}
%%%%%%%%%%%%%%%%%%%%%%%%%%%%%%%%%%%%%%%%%%%%%%%%%%%%%%%%%5
%%%%%%     sSLASH
\newsavebox{\sSLASH}
\sbox{\sSLASH}{$s$\hspace{-1.9mm}/}
\newcommand{\sS}{\usebox{\sSLASH}}
%%%%%%%%%%%%%%%%%%%%%%%%%%%%%%%%%%%%%%%%%%%%%%%%%%%%%%%%%5
%%%%%%     KSLASH
\newsavebox{\KSLASH}
\sbox{\KSLASH}{$k$\hspace{-1.8mm}/}
\newcommand{\KS}{\usebox{\KSLASH}}
%%%%%%%%%%%%%%%%%%%%%%%%%%%%%%%%%%%%%%%%%%%%%%%%%%%%%%%%%5
%%%%%%     LSLASH
\newsavebox{\LSLASH}
\sbox{\LSLASH}{$\ell$\hspace{-1.8mm}/}
\newcommand{\LS}{\usebox{\LSLASH}}
%%%%%%%%%%%%%%%%%%%%%%%%%%%%%%%%%%%%%%%%%%%%%%%%%%%%%%%%%5
%%%%%%     QSLASH
\newsavebox{\QSLASH}
\sbox{\QSLASH}{$q$\hspace{-1.8mm}/}
\newcommand{\QS}{\usebox{\QSLASH}}
%%%%%%%%%%%%%%%%%%%%%%%%%%%%%%%%%%%%%%%%%%%%%%%%%%%%%%%%%5
%%%%%%     DSLASH
\newsavebox{\DSLASH}
\sbox{\DSLASH}{$D$\hspace{-2.5mm}/}
\newcommand{\DS}{\usebox{\DSLASH}}
%%%%%%%%%%%%%%%%%%%%%%%%%%%%%%%%%%%%%%%%%%%%%%%%%%%%%%%%%5
%%%%%%     DSLASH
\newsavebox{\DbfSLASH}
\sbox{\DbfSLASH}{${\mathbf D}$\hspace{-2.8mm}/}
\newcommand{\DBFS}{\usebox{\DbfSLASH}}
%%%%%%%%%%%%%%%%%%%%%%%%%%%%%%%%%%%%%%%%%%%%%%%%%%%%%%%%%5
%%%%%   DELVECRIGHT
\newsavebox{\DELVECRIGHT}
\sbox{\DELVECRIGHT}{$\stackrel{\rightarrow}{\partial}$}
\newcommand{\PARVECR}{\usebox{\DELVECRIGHT}}

\title{\bf On the Charged Higgs Bosons Effects in the Top Quark Decays}
\author{ \bf M. Mohammadi
Najafabadi\thanks{email: mojtaba.mohammadi.najafabadi@cern.ch}\\ 
\\
{\it Department of Physics, Sharif University of Technology}\\
{\it P.O. Box 11365-9161, Tehran-Iran} \\
 and\\
{\it Institute for Studies in Theoretical Physics and Mathematics (IPM)}\\
{\it School of Physics, P.O. Box 19395-5531, Tehran-Iran} \\}

\date{}
\maketitle

\begin{abstract}
The collider experiments at the Tevatron and the LHC 
provide us the possibility of probing the existence of
a light charged Higgs boson. 
In this paper we study semi-leptonic decay of a polarized top quark
via the charged Higgs boson
($t(\uparrow)\rightarrow H^{+}b \rightarrow l^{+}\nu_{l}b$). 
It is shown that  
the asymmetry or spin correlation coefficient of the charged lepton
depends on the $\tan\beta$ and $M_{H^{\pm}}$ and is quite different from the 
Standard Model. 
This sensitivity of asymmetry to $\tan\beta$ and $M_{H^{\pm}}$ could be utilized
in the experimental searches for
the light charged Higgs for separation of the signal from backgrounds
in the $t\bar{t}\rightarrow H^{+}W^{-}b\bar{b}\rightarrow\tau^{+}\nu_{\tau}\l^{-}\bar{\nu_{\l}}b\bar{b}$ 
(and vice versa). It might be useful 
for obtaining better bounds in the ($\tan\beta,M_{H^{\pm}}$) plane too.
\end{abstract}

\vspace{4cm}
\hspace{0.8cm}

\newpage
\setcounter{page}{1}
%\tableofcontents
%\newpage
%\setcounter{page}{1}
\section{Introduction}
\setcounter{section}{1} 
The Standard Model of the particle physics has 
been found to be in a good agreement with the 
experimental observations with almost a high precision.
However, except for the Higgs boson of the Standard Model
which has not been observed yet, there are some 
other problems which directs physicists to a beyond Standard Model
theory. Some of the problems are mentioned below:
Gravitational interactions are not included in the Standard Model, 
within the Standard Model there is no candidate for the dark matter and  
there are some problems with the divergence of Higgs mass.

Presently, the Minimal Supersymmetric Standard Model (MSSM) is the 
main candidate for the unified theory beyond the Standard Model \cite{susyprimer}.  
The MSSM contains a two-Higgs doublets sector in which one doublet 
couples to the up-type quarks and neutrinos and the
other doublet couples to the down type quarks and charged leptons \cite{susyprimer,higgshunter}.

Search for Supersymmetry in the nature is one of the main
tasks of the experiments at colliders and non-accelerator experiments.
However, there are still no any direct indications on the existence of
Supersymmetry satisfying all theoretical and experimental requirements.
The scale of supersymmetry breaking might be around
1 TeV which allows the Tevatron and the LHC to explore supersymmetry.
The MSSM  has been predicted the existence of the charged Higgs boson.
There have been many searches for the charged Higgs boson 
at the LEP and the Tevatron and search for charged Higgs 
is one of the main tasks at the LHC \cite{Tev,LEP,leshouches,CMS}.
The production of $t\bar{t}$ pairs, with theoretical production cross section
of 6.7 pb (at the Tevatron) and 830 pb (at the LHC), may offer
a large source of charged Higgs production.
If kinematically allowed, the top quark can decay to 
$H^{+}b$, competing with the SM decay mode $t\rightarrow W^{+}b$ \cite{higgshunter}.
This mechanism not only provide a large production rate, in particular at the
LHC, but also a clean signature for the light charged Higgs \cite{Tev,hashemi}.
The results of direct and indirect searches 
for the charged Higgs at the LEP and the Tevatron 
excluded the region of $m_{H^{\pm}}<80-90$ GeV/c$^{2}$ 
(assuming $H^{+}\rightarrow\tau^{+}\nu_{\tau}$ and $H^{+}\rightarrow c \bar{s}$)
and the very low values of $\tan\beta$ ($\tan\beta < 1$) \cite{Tev,PDG}.

In this paper, we study the dependency of the 
{\it asymmetry} or {\it spin correlation coefficient} of the
charged lepton produced from a charged Higgs in the decay of 
a polarized top quark on the mass of the charged Higgs and $\tan\beta$.
Asymmetry of the charged lepton is strongly dependent on 
$m_{H^{\pm}}$ and $\tan\beta$ and could be used in separation
of the signal from backgrounds in the experimental searches
and for extracting some bounds 
on the ($\tan\beta,M_{H^{\pm}}$) plane.

This paper is organized as follows: Section 2 gives a 
brief description of the top quark spin correlations at hadron colliders. Section 3 is dedicated to
present the results of $t(\uparrow)\rightarrow H^{+}b\rightarrow l^{+}\nu_{l}b$.
Section 4 concludes the paper.

\section{The top spin correlations at hadron colliders}

The top quark is the heaviest fermion in the Standard Model with 
the mass of ~175 GeV/c$^{2}$. Such a large mass leads to 
a very short life time ($4\times 10^{-25}$s) which is one order of magnitude 
smaller than the QCD hadronization time scale ($3\times 10^{-24}$s).
Therefore, unlike the
other quarks it decays before hadronization and its spin 
information is transferred to its final products.
The dominant SM decay chain is $t\rightarrow W^{+}b\rightarrow l^{+}\nu_{l}b,q\bar{q}'b$.

The angular distributions of the final products of a spin-up top
quark are simply linear in the cosine of the decay angles:
\begin{eqnarray}
\frac{1}{\Gamma}\frac{d\Gamma}{d\cos\theta_{i}} = \frac{1}{2}\left(1+\alpha_{i}\cos\theta_{i}\right),
\end{eqnarray}
where $\theta_{i}$ is the angle between the {\it i}th decay product and
the spin vector of the top quark in the rest frame of the top quark
and $\alpha_{i}$ is called {\it asymmetry} or {\it spin correlation coefficient}.
The charged lepton and the down type quarks are maximally correlated 
with the top quark spin with $\alpha_{l,d} = 1$ \cite{Beneke,Mahlon}.

It is well known that top quarks produced at hadron colliders are scarcely polarized. As a result,
the correlations between top spin and anti-top spin in the $t\bar{t}$ is considered \cite{Beneke,Brandenburg}.
In order to illustrate we consider the dileptonic decay of the $t\bar{t}$ events in hadron colliders:
\begin{eqnarray}
\bf {PP,P\bar{P}} \rightarrow t\bar{t} + X \rightarrow l^{+}{l^{\prime-}} + X,
\end{eqnarray}
The double differential angular distribution of the leptons coming form the top and anti-top
in the ordinary SM is \cite{Beneke,Brandenburg}:
\begin{eqnarray}\label{spintt}
\frac{1}{\sigma}\frac{d^{2}\sigma}{d\cos\theta_{l^{+}}d\cos\theta_{l'^{-}}} = \frac{1}{4}(1+
\kappa \cos\theta_{l^{+}}\cos\theta_{l'^{-}})~,~\kappa = \alpha_{l^{+}}\alpha_{l'^{-}}
\times\frac{N_{\|}-N_{\times}}{N_{\|}+N_{\times}}.
\end{eqnarray}
where $\theta_{l^{+}}$$(\theta_{l'^{-}})$ is the angle between the direction of the $l^{+}(l'^{-})$  
in the rest frame of the $t(\bar{t})$ and the $t(\bar{t})$ direction in the $t\bar{t}$ center of mass.
$N_{\|}$ is the number of top
pair events where both quarks have spin up or spin down
and $N_{\times}$ is the number of top pair events where one quark is spin up and the other is spin down.
Eq.(\ref{spintt}) shows the strong dependence of the experimental observable, $\kappa$, to the asymmetry 
 ($\alpha_{l}$).
For the $t\bar{t}$ production at the Tevatron and the LHC,
the SM predicts $(N_{\|}-N_{\times})/(N_{\|}+N_{\times}) = 0.88,0.33$, respectively. 
In \cite{CMS}, it has been shown that $\kappa$ will be measured 
very precisely at the LHC.

\section{The decay of $t(\uparrow)\rightarrow H^{+}b\rightarrow l^{+}\nu_{l}b$}

\begin{figure}[ht]
\begin{center}
\epsfig{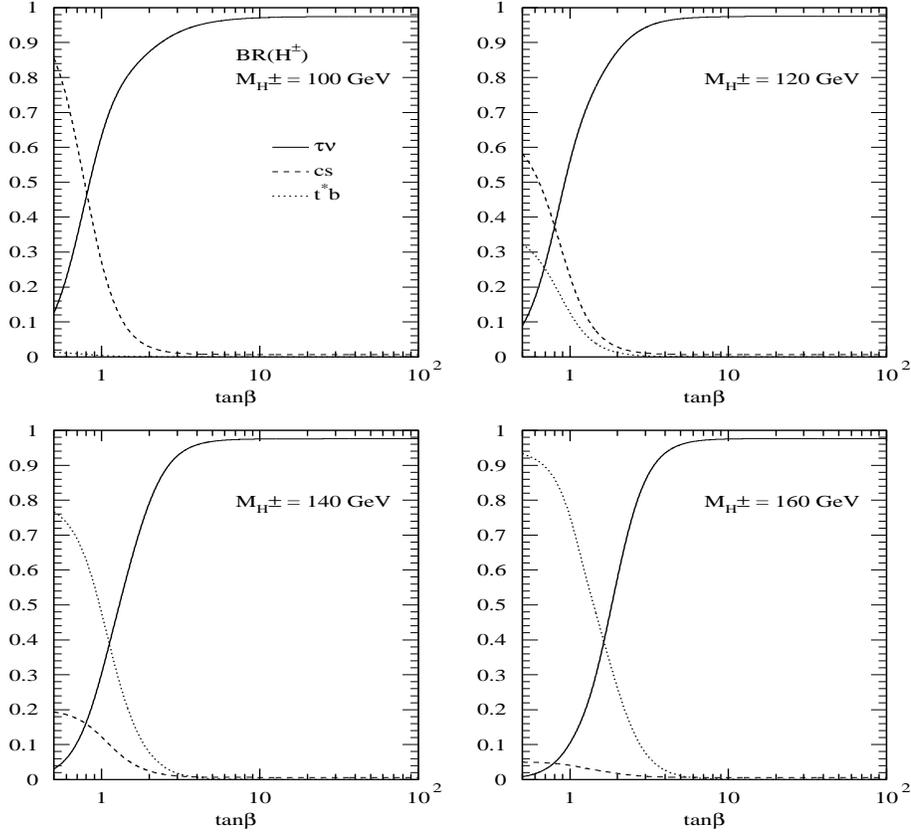}
 \caption{The branching fractions of the charged Higgs boson as a function of $\tan\beta$
 for four values of the charged Higgs mass \cite{tevatron}.}
    \label{fig:br}
\end{center}
\end{figure}

The Lagrangians describing the $H^{+}tb$ and $H^{+}l\nu$ vertices in the MSSM read
as follows \cite{higgshunter}:
\begin{eqnarray}\label{lag}
\mathcal{L}_{Htb} &=& \frac{gV_{tb}}{\sqrt{2}m_{W}}H^{+}\bar{t}\left(m_{t}\cot\beta P_{L}+
m_{b}\tan\beta P_{R} \right)b + h.c. , \nonumber \\
\mathcal{L}_{Hl\nu} &=& \frac{g m_{l}\tan\beta}{\sqrt{2}m_{W}}H^{+}\bar{\nu_{l}}
 P_{R}l + h.c. , 
\end{eqnarray}
where $P_{L,R} = \frac{1\mp \gamma_{5}}{2}$ are the chiral projection operators,
V$_{tb}$ is the CKM element which is set to one in the rest of this work.
Because charged Higgs couplings are proportional to 
fermion masses, the decays to the third generation of the quarks
and leptons are dominant. In particular, for the light charged Higgs 
coming from a top quark when $\tan\beta$ is large enough ($\tan\beta \geq 4-5$),
 $BR(H^{+}\rightarrow\tau^{+}\nu_{\tau})\simeq 1$, which is an apparent breakdown of 
the $e-\mu-\tau$ universality \cite{higgshunter}.
Fig.\ref{fig:br} shows the branching fractions of the charged Higgs boson as a function of $\tan\beta$
corresponding to four different values of the charged Higgs boson mass \cite{tevatron}.

Using the Lagrangians from Eq.(\ref{lag}), the squared matrix element for the 
reaction $ t(p_{1})\rightarrow H^{+}(q) + b(p_{2}) \rightarrow
l^{+}(k_{1})+\nu_{l}(k_{2})+b(p_{2})$ has the following form:
\begin{eqnarray}\label{mtx}
\overline{|\mathcal{M}|}^{2} &=& 
\frac{16G_{F}^{2}\tan^{2}\beta m_{\tau}^{2}}{(q^{2}-m_{H}^{2})^{2}+\Gamma_{H}^{2}m_{H}^{2}}(k_{1}.k_{2})\nonumber\\
&\times& \left(2m_{t}^{2}m_{b}^{2}+p_{1}.p_{2}(m_{t}^{2}\cot^{2}\beta+m_{b}^{2}\tan^{2}\beta)
+m_{t}p_{2}.s(m_{t}^{2}\cot^{2}\beta-m_{b}^{2}\tan^{2}\beta)\right),
\end{eqnarray}
where $G_{F}=\frac{g^{2}}{8m^{2}_{W}}$ is the Fermi constant, $s^{\mu} =
(0,\vec{s})$ is the polarization four-vector of the top quark, $m_{H}$ and $\Gamma_{H}$ are
the charged Higgs mass and the full width of the charged Higgs, respectively.

After following the same method as used in \cite{Kuhn2,Kuhn3} and  
performing some algebra
the double differential $(x-\theta_{l})$, energy-angular, distributions in the point-like 
four fermion limit is:
\begin{eqnarray}\label{DWidth}
\frac{d^{2}\Gamma}{dx~d\cos\theta_{l}}&=&\frac{G^{2}_{F}m^{3}_{t}m^{2}_{\tau}m_{H}}{128\pi^{2}\Gamma_{H}}
\times\left(A+B(x)\cos\theta_{l}\right)\nonumber\\
A &=& 4r_{2}+(1+r_{2}-r_{1})(\cot^{2}\beta+r_{2}\tan^{2}\beta),\nonumber\\
B(x) &=& (r_{1}-r_{2}+1-\frac{2r_{2}}{x})(\cot^{2}\beta-r_{2}\tan^{2}\beta),\nonumber\\
r_{1} &=& \frac{m_{H}^{2}}{m_{t}^{2}}~ ,~ r_{2} = \frac{m_{b}^{2}}{m_{t}^{2}}~,~\frac{2m_{l}}{m_{t}}\leq x \leq 1,
\end{eqnarray}
where $ x = \frac{2E_{l}}{m_{t}}$, $E_{l}$ is the energy of the charged lepton,
and $\theta_{l}$ is the angle between 
the spin of the top quark and the momentum of the lepton in the rest
frame of the top quark. In deriving Eq.(\ref{DWidth}), the narrow width 
approximation of the charged Higgs has been  used:
\begin{eqnarray}
\frac{1}{(q^{2}-m_{H}^{2})^{2}+\Gamma_{H}^{2}m_{H}^{2}}\rightarrow \frac{\pi}{\Gamma_{H}m_{H}}\delta(q^{2}-m^{2}_{H}).
\end{eqnarray}
It is more useful to express the Eq.(\ref{DWidth}) in the following way:
\begin{eqnarray}\label{Dwidth1}
\frac{d^{2}\Gamma}{dx~d\cos\theta_{l}}=\frac{d\Gamma}{dx}\times\frac{1}{2}\left(1+\alpha_{l}(x)\cos\theta_{l}\right),
\end{eqnarray}
where $\alpha_{l}(x)$, {\it correlation coefficient} or {\it asymmetry}, is:
\begin{eqnarray}
\alpha_{l}(x) = \frac{B(x)}{A}.
\end{eqnarray}
\begin{figure}[ht]
\begin{center}
\epsfig{file=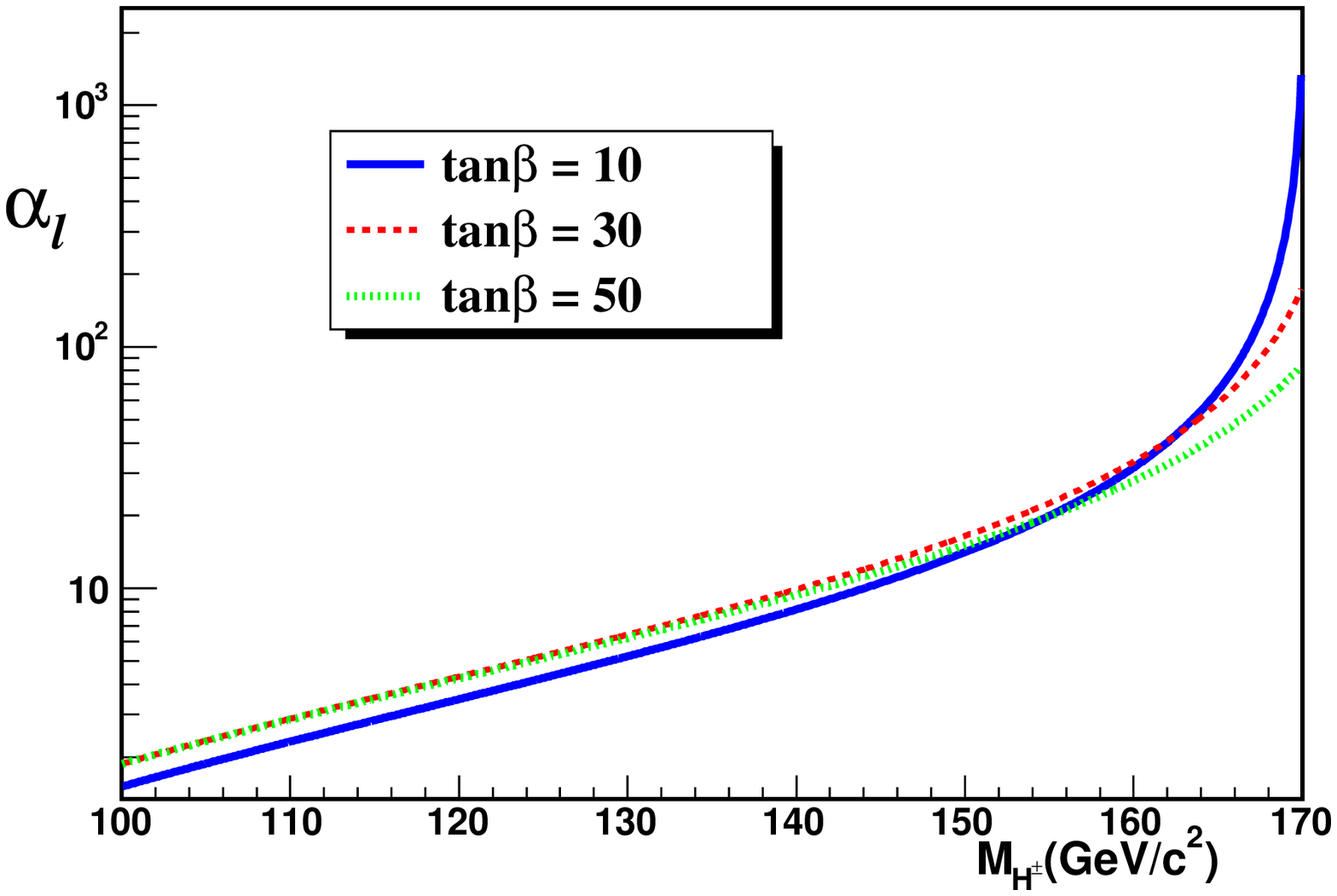,height=7cm,width=7cm}
\epsfig{file=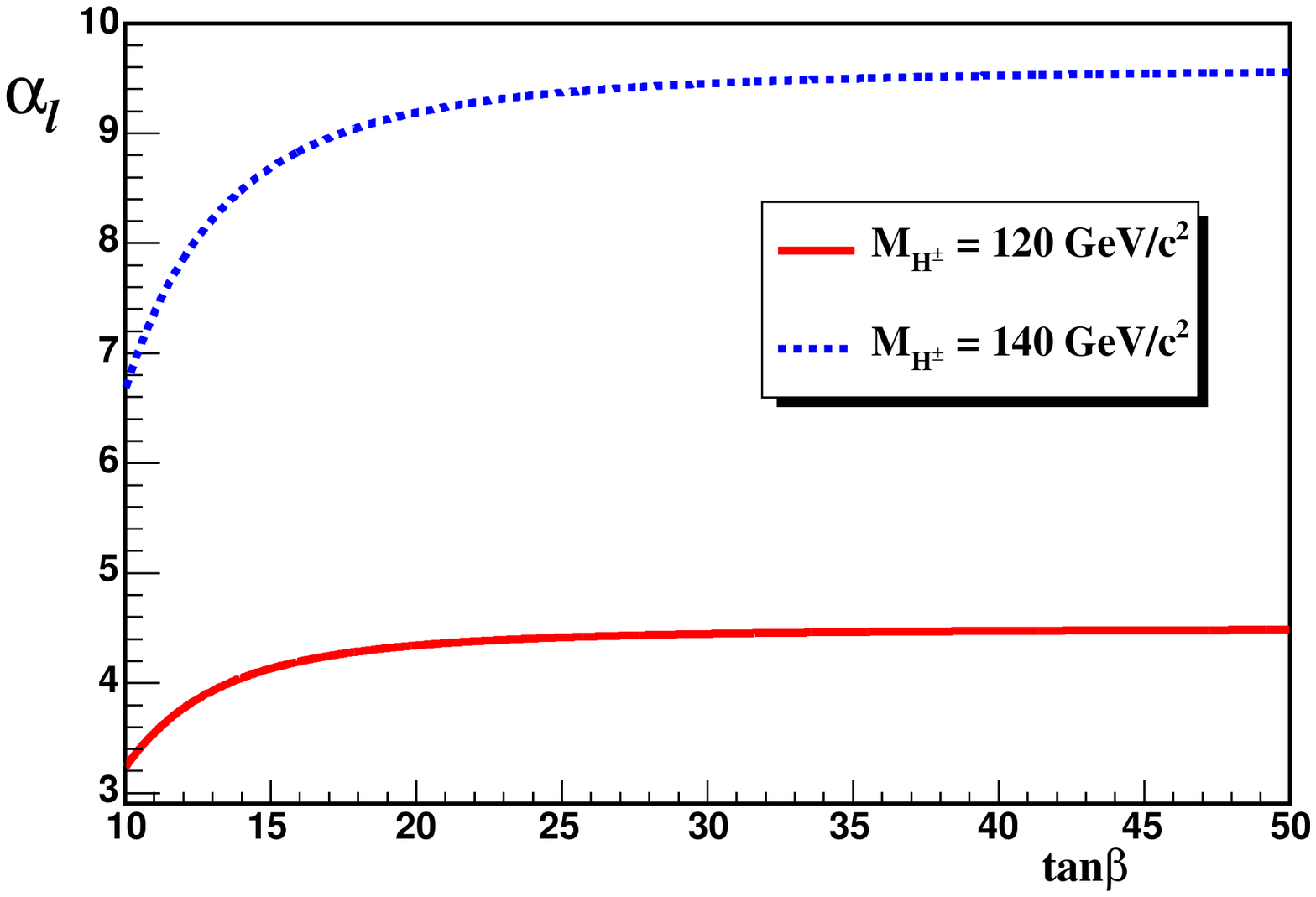,height=7cm,width=7cm}
 \caption{ Left: Asymmetry as a function of the charged Higgs mass. Right: 
   Asymmetry as a function of $\tan\beta$.}
    \label{fig:CorrCoef}
\end{center}
\end{figure}

In the ordinary standard model and in the limit of vanishing the lepton masses, 
$\alpha_{l}$ is independent of $x$ and is equal to one \cite{Mahlon}. As a result of Eq.(\ref{Dwidth1}), 
the angular distribution of the final 
charged lepton in the decay of a polarized top via a charged Higgs 
is linear in cosine of the decay angle, the same form as
the ordinary Standard Model, except for the case that $\tan\beta = \sqrt\frac{m_{t}}{m_{b}}$.
Therefore, an interesting observation is that
when $\tan\beta = \sqrt\frac{m_{t}}{m_{b}}\simeq 6.0$ 
the angular distribution of the final charged lepton is not linear in $\cos\theta_{l}$ 
and is constant. 
\begin{figure}[ht]
\begin{center}
\epsfig{file=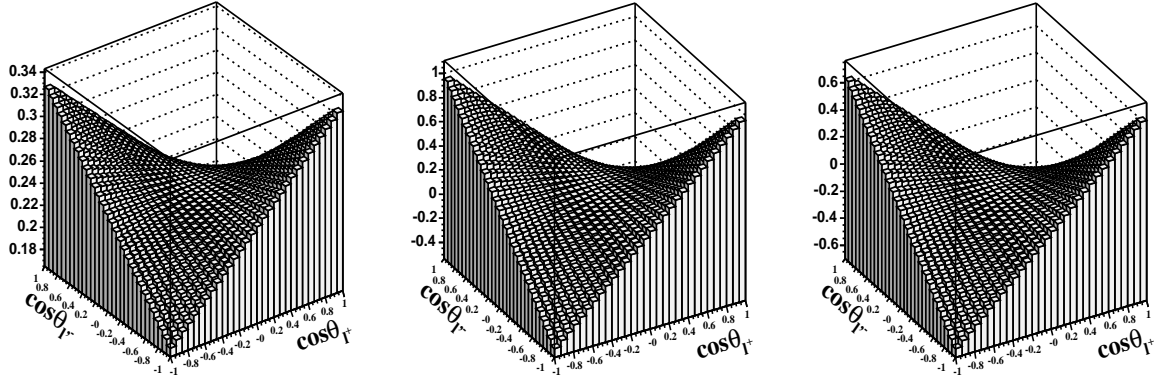,height=6cm,width=16cm}
 \caption{ The distribution of $\frac{1}{\sigma}\frac{d^{2}\sigma}{d\cos\theta_{l^{+}}d\cos\theta_{l'^{-}}}$  for
   $t\bar{t}\rightarrow W^{+}W^{-}b\bar{b}\rightarrow l^{+}l'^{-}b\bar{b}$ (Left),
   $t\bar{t}\rightarrow H^{+}W^{-}b\bar{b}\rightarrow\tau^{+}\nu_{\tau}\l^{-}\bar{\nu_{\l}}b\bar{b}$ (Middle) with $\tan\beta = 20$ and $m_{H} = 140$, The difference of the Middle and the Left distributions (Right). 
  }
    \label{fig:SPIN}
\end{center}
\end{figure}

According to Eq.(\ref{DWidth}) and Eq.(\ref{Dwidth1}) the
asymmetry, $\alpha_{l}$, depends on the charged Higgs mass and $\tan\beta$.
After integration over $x$ in Eq.(\ref{DWidth}) $\alpha_{l}$
can be extracted,
Fig.\ref{fig:CorrCoef} shows the 
dependence of the $\alpha_{l}$ on the $m_{H}$ and $\tan\beta$.
It is clear that there is a significant difference from the Standard Model.
According to the recent searches for the light charged Higgs
in $t\bar{t}\rightarrow H^{\pm}W^{\mp}b\bar{b}\rightarrow\tau^{\pm}\nu_{\tau}\l^{\mp}\bar{\nu_{\l}}b\bar{b}$ 
\cite{Tev,hashemi}, the most dangerous background is $t\bar{t}\rightarrow W^{+}W^{-}b\bar{b}$ (with 
at least one $W$ decaying leptonically).

Fig.\ref{fig:SPIN}, shows the distributions of the Eq.(\ref{spintt})
for the Standard Model case in the dileptonic decay mode (left), for the $t\bar{t}\rightarrow 
H^{+}W^{-}b\bar{b}\rightarrow\tau^{+}\nu_{\tau}\l^{-}\bar{\nu_{\l}}b\bar{b}$ with 
$\tan\beta = 20$ and $m_{H} = 140$ (the middle distribution) and 
the distribution in the right side shows the difference of the 
middle and the left distributions. According to the right distribution
in Fig.\ref{fig:SPIN}, in the case of $t\bar{t}\rightarrow 
H^{+}W^{-}b\bar{b}\rightarrow\tau^{+}\nu_{\tau}\l^{-}\bar{\nu_{\l}}b\bar{b}$
events prefer to be in the regions of $\cos\theta_{l^{+}} > 0$ and $\cos\theta_{l'^{-}} > 0$
rather than the Standard Model case.
Therefore, constructing the distribution of Eq.(\ref{spintt}) and applying 
some optimal cuts on the $\cos\theta_{l^{+}}$ and $\cos\theta_{l'^{-}}$
might suppress the $t\bar{t}\rightarrow W^{+}W^{-}b\bar{b}\rightarrow l^{+}l'^{-}b\bar{b}$ background and leads to 
a better significance and cleaner signal in the searches for the charged Higgs produced in the $t\bar{t}$.
Measurement of the $\kappa$ in Eq.(\ref{spintt}) can help to achieve 
some new bounds on the ($\tan\beta,M_{H^{\pm}}$) plane.

\section{Conclusion}
The angular distribution of the produced charged lepton ($\tau^{\pm}$)
in the decay of a polarized top quark via 
a charged Higgs was studied. The same as the Standard Model, this distribution 
is linear in cosine of the angle between the spin of the top 
and the momentum of the charged lepton except for the 
case of $\tan\beta = \sqrt \frac{m_{t}}{m_{b}}$.
The asymmetry, which is somehow experimentally observable,
is dependent on the $m_{H^{\pm}}$ and $\tan\beta$  
and differs from  
the Standard Model value significantly. In the study of the
light charged Higgs generated from $t\bar{t}$ at the hadron collider,
in particular at the LHC with a lot of 
$t\bar{t}$ events (eight millions per year), the $t\bar{t}$ spin correlation can
help seriously to suppress backgrounds.
It also might be useful to obtain bounds on the ($\tan\beta,M_{H^{\pm}}$) plane. 

\section{Acknowledgment}
The author thanks to S. Paktinat for the useful comments.


\begin{thebibliography}{99}
\bibitem{susyprimer} S.P. Martin, {\it A Supersymmetry Primer}, [hep-ph/9709356].
\bibitem{higgshunter} J.F. Gunion, H.E. Haber, G.L. Kane and S. Dawson, 
{\it The Higgs Hunter's Guide}, (Addison-Wesley, New York, 1990).
\bibitem{Tev} A. Abulencia et al. (CDF Collaboration), {\it Search for charged Higgs
bosons from top quark decays in $p\bar{p}$ collisions at $\sqrt s= 1.96$ TeV},
Phys. Rev. Lett. \textbf{96} (2006) 042003 [hep-ex/0510065].
\bibitem{LEP} J. Abdallah, et al. (DELPHI Collaboration), 
{\it Search for Charged Higgs Bosons at LEP in General Two Higgs Doublet Models},
Eur.Phys.J. \textbf{C34} (2004) 399.
\bibitem{leshouches} K.A. Assamagan, {\it Proceedings 3rd Les Houches Workshop: Physics at TeV Colliders},
[hep-ph/0406152].
\bibitem{CMS} {\it CMS Physics Technical Design Reports, Volume II: Physics Performance}, CERN/LHCC 2006-021.
\bibitem{hashemi} M. Baarmand, M. Hashemi and A. Nikitenko, {\it Light charged Higgs discovery 
potential of CMS in the $H^{\pm}\rightarrow\tau\nu_{\tau}$ decay with single lepton trigger},
J. Phys. G: Nucl. Part. Phys. \textbf{32} (2006) N21-N40.
\bibitem{PDG} S. Eidelman et al. (Particle Data Group), Phys. Lett. \textbf{B592}, (2004) 1.
\bibitem{Beneke} M. Beneke et al., {\it  Top Quark Physics}, [hep-ph/0003033].
\bibitem{Mahlon} G. Mahlon, S. Parke, 
{\it Angular correlations in top quark pair production and decay at hadron colliders},
Phys. Rev. \textbf{D53} (1995) 4886.
\bibitem{Brandenburg} W. Bernreuther, A. Brandenburg, Z. S. Si, P. Uwer, 
{\it Top quark spin correlations at hadron colliders: Predictions at next-to-leading order QCD},
Phys. Rev. Lett. \textbf{87} (2001) 242002.
\bibitem{tevatron}  M. Carena, et al. {\it Report of the Higgs Working Group of the Tevatron Run 2 SUSY/Higgs Workshop},
[hep-ph/0010338].
\bibitem{Kuhn2} A.~Czarnecki, M.~Jezabek, J.H.~Kuhn,
{\it Lepton Spectra from Decays of Polarized Top Quarks}, 
Nucl. Phys. \textbf{B351} (1991) 70.
\bibitem{Kuhn3} M.~Jezabek, J.H.~Kuhn, {\it Distributions of Leptons in Decays
of Polarised Heavy Quark}, Nucl. Phys. \textbf{B427} (1994) 3.
\end{thebibliography}
\end{document}